\documentstyle{laa}

\begin{document}

\thesaurus{11 (11.03.1; 11.03.4 Abell 2390; 12.03.3; 12.04.1; 12.07.1;
 13.25.2)}
123456789 123456789 123456789 123456789 123456789 123456789 123456789 123456789
\title{X-ray analysis and matter distribution in the lens-cluster Abell 2390 }

\author{
M. Pierre\inst{1}
\and J.F. Le Borgne\inst{2}
\and G. Soucail\inst{2}
\and J.P. Kneib\inst{2,3}
}

\offprints{M. Pierre}

\institute{CEA Saclay, Service d'Astrophysique, F-91191 Gif sur Yvette
 \and
 Observatoire Midi-Pyr\'en\'ees, Laboratoire d'Astrophysique de Toulouse,
 URA 285, 14 Avenue E. Belin, F-31400 Toulouse
 \and
 Institute of Astronomy, Madingley Road, Cambridge CB3 0HA, UK
 }

\date{Received ?? , Accepted ???}
\maketitle

\begin{abstract}
We present a deep X-ray ROSAT/HRI observation of the rich cluster-lens Abell
2390 and the analysis of the gas and matter distribution in the cluster.
The  X-ray centroid coincides with the location of the central cD galaxy,
and the core appears to be  dominated by a massive cooling flow, one of the
strongest ones ever studied. On larger scales, the X-ray distribution is
elliptical and locally distorted by a significant ``sub-structure" located
some 100 $h^{-1}$ kpc from the center. The  mass of this sub-structure
is estimated to be $\sim$ 1/20 of the total cluster mass and appears to
be associated with an enhancement in the galaxy distribution,
in the region where the giant
``straight arc" and several arclets are observed. A simple lens model based
on the geometrical properties of the X-ray gas shows that this sub-structure
allows to explain in a simple manner the peculiar shape of the arc and
puts strong constraints on the total mass distribution in the central
part of this cluster. The derived M/L in the central part of the cluster is
found to be around 250 in solar units, a rather common value compared with
other cluster-lenses. Moreover, the link between the matter distribution
derived from the lens model and the density profile issued from the
X-ray map analysis is used to infer the temperature profile in the cluster.
This profile shows a significant decrease near the center, in agreement
with the existence of the cooling flow. Both X-ray and lensing constraints are
tied together to give a consistent picture of the dynamics of the central
part of A2390.

\end{abstract}

\section {Introduction}
The problem of mass determination in clusters of galaxies is still a debated
question although some significant advances have been recently achieved,
often related to the improvement of the observing capabilities.
So far, three independent methods are currently used to determine the mass
 distribution  in galaxy clusters,
each of them having its own limitations and uncertainties. First,
the galaxy distribution combined with the velocity dispersion gives the
virial mass. Some more refined analysis of the velocity distribution
are now explored (Merritt 1987; Merritt \& Saha 1994;
Merritt \& Tremblay 1994). Secondly, the X-ray emission of the optically thin
intracluster gas has long been thought to be the ideal tracer of the mass.
Indeed, the collisional timescale of the ICM being much shorter than the
Hubble time, one generally considers that the free-free emitting gas is in
hydrostatic equilibrium in the cluster potential.
Thus, the X-ray surface brightness of the cluster is simply related to the
kinetic temperature of the gas and to the square of the electron density.
The main difficulty arises while deprojecting the 2D X-ray map into
a density profile, as this operation implicitly assumes spherical symmetry
and requires in principle the knowledge  of the temperature distribution.

The third approach is to use the gravitational
lensing effect observed through rich and medium-distant clusters producing
giant luminous arcs and arclets (Fort \& Mellier 1994).
The modelling of multiple arcs
is a powerful tool to infer the central mass distribution and to
constrain the core radius of the dark matter
(e.g. Mellier et al. 1993; Kneib et al. 1993). From the analysis of several
clusters, the estimated core radius seems about twice smaller than
expected from the standard analysis of X-ray images (Kneib et al. 1995; Smail
et al. 1995a); this  discrepancy is not well understood yet,
although probably due to the fact that,  until very recently,
the X-ray detector PSF  has never been taken into account,
so that published core radii are actually convolved values.
Recent works (Loeb \& Mao 1994; Miralda-Escud\'e \& Babul 1995;
Daines et al. 1995; Kneib et al. 1995; Allen et al. 1995)
combine the X-ray and lensing analysis
to study the dynamics of clusters. The morphology of the mass distribution
infered from the X-ray and lensing analyses agree and the mass centroid
coincides with the brightest cluster galaxy
(usually a cD or giant elliptical); the mass distribution having same
orientation and ellipticity
as the halo of the central galaxy. However, in some clusters such
as A1689 and A2218, a discrepancy by a factor two in the mass estimates
are noticed (Miralda-Escud\'e \& Babul 1995). It can be explained by
the fact that the cluster is merging and has not reach the state of
hydrostatic equilibrium yet. On the contrary,
in the cluster-lens PKS0745 (Allen et al. 1995) lensing and X-ray
constraints do agree when a multiphase-model of the IGM (based on ROSAT/HRI
and ASCA data) is used.

In this context, Abell 2390 (z=0.232) is an attractive object to  study in
many respects: first, a ``straight'' giant gravitational
arc has been detected by Pell\'o et al. (1991, hereafter Paper I)
with its redshift measurement
(z=0.913) and the identification of several arclets, indicating that the
cluster may be massive and concentrated in its center. Moreover, the galaxy
distribution presents an elongated shape, and the velocity dispersion
$\sigma= 2112 ^{+274}_{-197}$ km s$^{-1}$ is quite large (Le Borgne et al.
1991, Paper II). This value may be overestimated by selection
biases in the sample, and could be slightly reduced.
Finally, strong evidences for a massive cooling flow arise from the detection
of bright optical filaments in the envelope of the central cD galaxy,
emitting low excitation emission lines, and a powerful unresolved
radio source (Soucail et al. 1995, Paper IV).
Einstein observation revealed a strong X-ray source  (Ulmer et al. 1986),
which when computed with the correct redshift corresponds to
$L_X = 3.85 \, 10^{44} \, h^{-2}$ ergs s$^{-1}$ in the range 0.7--3.5
keV. This source
is significantly elongated, with a centrally condensed core (Mc Millan et al.
1989). From its high X-ray luminosity and  galaxy velocity
dispersion, one may infer a large cluster mass;
alternatively this could  indicate that the cluster  has not reached a
relaxed state yet, in agreement with its overall very elongated morphology.
This, together with the exceptional shape of the giant arc,
may suggest the existence of a peculiar underlying gravitational
potential as proposed by Kassiola et al. (1992).
But despite the numerous arclets observed, a univoque
modelling was not possible till now, because of the lack of lensing
constraints within the Einstein radius. In the classical hypothesis where
the hot emitting X-ray plasma is a much better tracer of the overall
mass than the galaxies - which seems justified because of its considerably
shorter mean free path - the gas distribution should be the most direct tool
to discriminate between potential models derived from optical
constraints alone. Thus, the need for a detailed mapping of the cluster
central parts has motivated the obtention of a deep ROSAT HRI
pointing which we analyse in this paper.

The paper is organised as follows. In Section 2, we present the X-ray and
optical observations, as well as a morphological analysis of the maps and
the significance of the different detected structures. Section 3 deals with the
analysis of the gas properties derived from the X-ray flux and profile.
In Section 4, the lensing properties of the cluster are examined in connection
with the constraints derived from the gas distribution. The mass profile
obtained from the lens modelling is then used to estimate the gas temperature
profile. Finally, some
conclusions on the matter distribution are given in Section
5. In all the paper we assume that $H_{0}=100 \, h$ km s$^{-1}$
Mpc$^{-1}$ ($0.5<h<1$), $\Omega_{0}=1$ and $\Lambda=0$,
which means that 1\arcsec\ scales as 2.33 $h^{-1}$ kpc at the
cluster redshift; $r$
and $R$ stand respectively for the 3D and projected radial distances
from the cluster center. Celestial coordinates are given in the J2000
reference system.

\section{ X-ray/optical morphological analysis}
\begin{figure*}
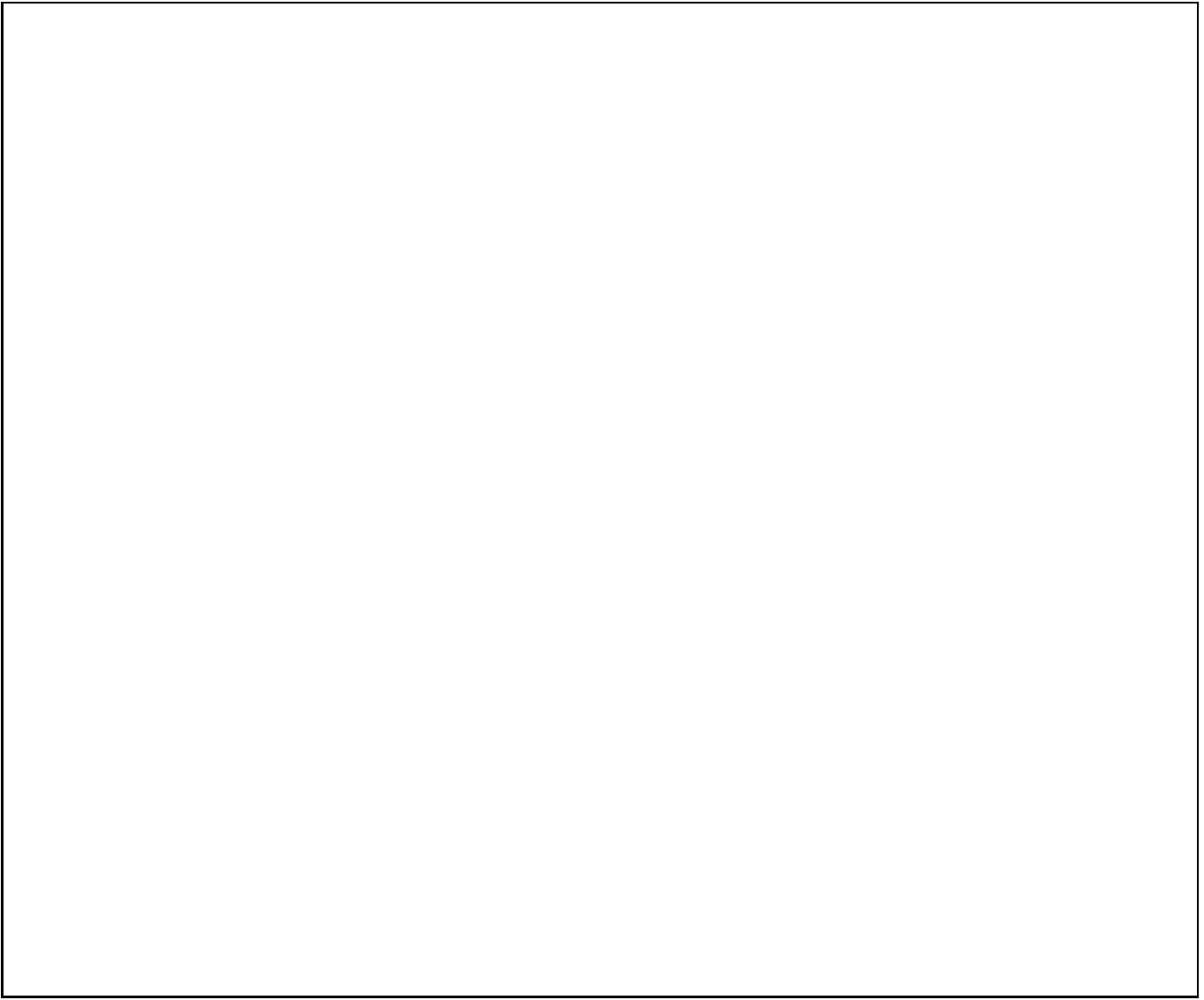

\picplace{15cm}
\caption [] {Deep R-band image of A2390 obtained at the KPNO 4m telescope.
The displayed field of view is 5\arcmin $\times$ 5\arcmin , centered on the
cD galaxy.
Overlaid are the ROSAT HRI contours equally separated in log space from
0.23 to 4.08 $\times 10^{-5}$ counts s$^{-1}$ arcsec$^{-2}$.
The X-ray image has been smoothed using a variable gaussian filter (see
text).}
\end{figure*}

\subsection{X-ray observations and optical re-centering}
Pointed ROSAT HRI observations were obtained between November 23-25 1992,
during a  single exposure totalling  27764 sec. All processing of the
X-ray data has been performed with the EXSAS/MIDAS package.
The photon file was binned into a 1\arcsec\ pixel image, and smoothed by a
gaussian filter having a  variable width function of the
pixel intensities: from 5\arcsec\
(nominal HRI FWHM) at the cluster center up to 20\arcsec\
at the periphery; this enables to reduce the photon noise in
the external isophotes, without loosing spatial resolution in  the brightest
central region. Extended emission
from the cluster  is detected beyond 6\arcmin\  (i.e. at
least 1 $h^{-1}$ Mpc).

Figure 1 shows the X-ray contours overlaid on the cluster optical image.
Absolute positioning of the X-ray image  was directly obtained from
the EXSAS image header,  but the alignment of the optical image requires
the precise knowledge of the cD coordinates which
have to be determined relatively  to surrounding known stars. The
coordinates of the central galaxy published in Paper II
have an uncertainty of about 2\arcsec\ so we tried to improve the
accuracy to better than 1\arcsec . The method we used consists in computing
 the transformation parameters between CCD coordinates and celestial
coordinates with a least square minimisation on
coordinates of stars of known equatorial coordinates measured on a CCD frame
(in practice the Thuan-Gunn r image from the
Isaac Newton Telescope described in Paper II).
Because we do not have unsaturated stars of known coordinates on the CCD
frames this was done in two steps:
\begin{itemize}
\item a) On a print of the Palomar Observatory Sky Survey (POSS) we
measured the position
of 22 stars of the HST Guide Star Catalogue in a field of about
15\arcmin $\times$15\arcmin\ and computed the transformation parameters
for the POSS print.
At this point the standard deviation of the residuals is 0.7\arcsec . These
parameters were then used to calculate the equatorial coordinates of 11
apparently stellar objects selected on a CCD frame where they appear
unsaturated.
\item b) The positions of these stellar objects were measured on the CCD frame.
Their equatorial coordinates computed in a) were used to calculate the
transformation parameters for the CCD frame. These parameters allowed to
calculate the equatorial coordinates of the center of the central galaxy
on the CCD frame. Although we did
not improve the accuracy significantly, this second step was necessary
because the position of the central galaxy was not well enough defined on
the POSS print while it could be defined to within half a pixel
(0.4\arcsec) on the CCD image.
\end{itemize}
The coordinates obtained for the central galaxy are:
$\alpha_{2000}$ = 21h 53m 36.76s, $\delta_{2000}$ = 17$^\circ$ 41\arcmin
42.9\arcsec.
The uncertainties are 0.04s in right ascension and 0.7\arcsec\ in
declination at a 99\% confidence level. These coordinates are slightly
different from those published in Paper II but they agree within uncertainties.

The X-ray emission is consistent with being centered on the cD  galaxy  within
1\arcsec\ (the brightest pixel on the filtered image is at
$\alpha_{2000}$= 21h 53m 36.67s, $\delta_{2000}$= 17$^\circ$  41\arcmin
43.7\arcsec ) provided that the absolute  ROSAT attitude solution
for this pointing is correct. The X-ray maximum appears
extremely concentrated and peaked, suggesting the presence of a massive
cooling flow, a first indication of the results discussed below.

\subsection{Morphological analysis of the X-ray 2D distribution}
\begin{figure}
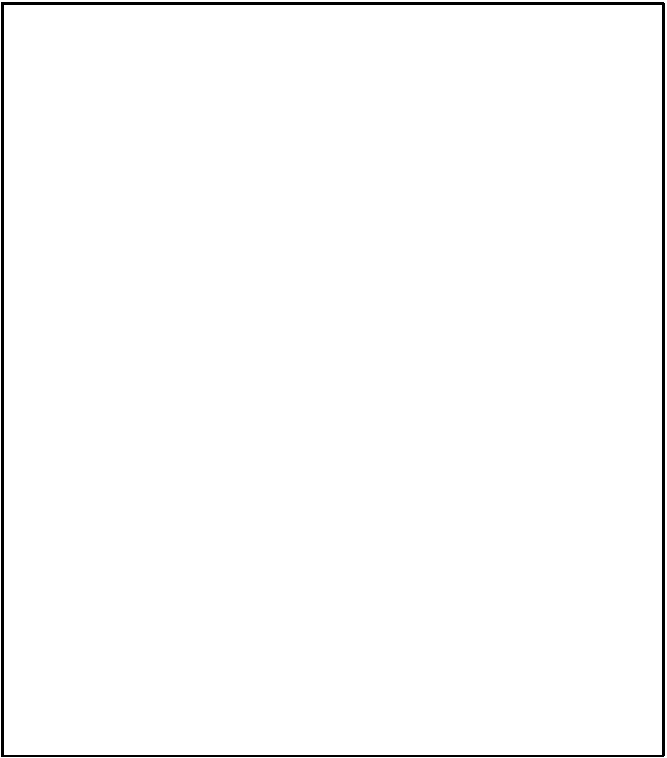

\picplace{10cm}
\caption [] {Elliptical modelling of the X-ray image as a function of radial
distance along the major axis. The squares correspond to the X-ray isophotes
fit (12\arcsec $<$ R $<$ 120\arcsec ) and the crosses to the cD envelope fit,
performed from the
R CCD image (2\arcsec $<$ R $<$ 20\arcsec ). The upper graph represents the
axis ratio distribution ($b/a$) and the lower one the
position angle distribution of the major axis, measured counterclockwise from
the North.
}
\end{figure}

On large scales, the X-ray image is strongly elliptical with an overall
position angle  comparable to the main cluster direction in the optical
(i.e. galaxy distribution and cD major axis, see below).
A conspicuous distortion of the isophotes due to an excess of emission
is located some 40\arcsec\ North-West from  the cD and another one
appears marginally on the opposite side of the cD (Fig. 1).
In order to study the 2D distribution of the gas and the departure from a
pure elliptical distribution, a detailed isophotal fit of the X-ray smoothed
map was done, using the ELLIPSE package in the IRAF/STSDAS environment
(Jedrzejewski 1987), as well as a wavelet analysis.

For the isophotal fit analysis, the two regions departing from a regular
elliptical isophote were masked before the fit.
Outside the central region, the fit was rather easy,  with an
increase of the ellipticity and a twist of the isophotes at large
radius (Fig. 2). Near the center, there is also a strong rotation of the
position angle of the isophotes and a significant decrease in the
ellipticity for $R<10\arcsec$. This may be due to some features at the
very center, possibly poorly resolved with our data (see below, wavelet
analysis) so that we did not consider residuals within the innermost
10\arcsec.
The elliptical model was subtracted to the initial X-ray map,
and the residuals overlayed on the optical image (Fig. 3).
In addition to the central residuals, three excess of emission are
clearly visible on this map, noted P1, P2 and S. P1 and P2 appear
unresolved while S is certainly extended. In order to estimate the
significance of each residual, we computed within each aperture
the number of counts in the elliptical model and derived a
signal-to-noise ratio of S/N = $N / \sqrt{N+N_{mod}}$.
Their main morphological properties are summarised in Table 1.

\begin{table}
\caption[ ]{Morphological properties of the residuals of the X-ray
elliptical fit. Columns 2 and 3 give the offsets in RA and DEC from the
cD galaxy, columns 4 gives the radial distance in kpc. In column 5, we
estimate the significance of the excess (see text for more details),
and in column 6 we  give the relative intensity, scaled to the total
cluster flux. Finally in column 7, we give the estimated X-ray luminosity,
assuming a 2 keV temperature at the cluster redshift.
}
\begin{flushleft}
\begin{tabular}{lllllll}
\noalign{\smallskip}
\hline
\noalign{\smallskip}
 & $\Delta \alpha$ & $\Delta \delta$ & Distance
& S/N & $F/F_{tot}$ &  Luminosity\\
 & (\arcsec) & (\arcsec) & ($h^{-1}$kpc) & & &
($h^{-2}$erg s$^{-1}$) \\
\noalign{\smallskip}
\hline
\noalign{\smallskip}
P1 & --17 & +9 & 44 & 1.1 & 3.4 $10^{-3}$ & 1.6 $10^{42}$ \\
P2 & --24 & --9 & 61 & 1.8 & 5.4 $10^{-3}$ & 2.6 $10^{42}$ \\
S & --37 & --25 & 105 & 2.6 & 9.0 $10^{-3}$ & 4.4 $10^{42}$ \\
\noalign{\smallskip}
\hline
\end{tabular}
\end{flushleft}
\end{table}

\begin{figure}
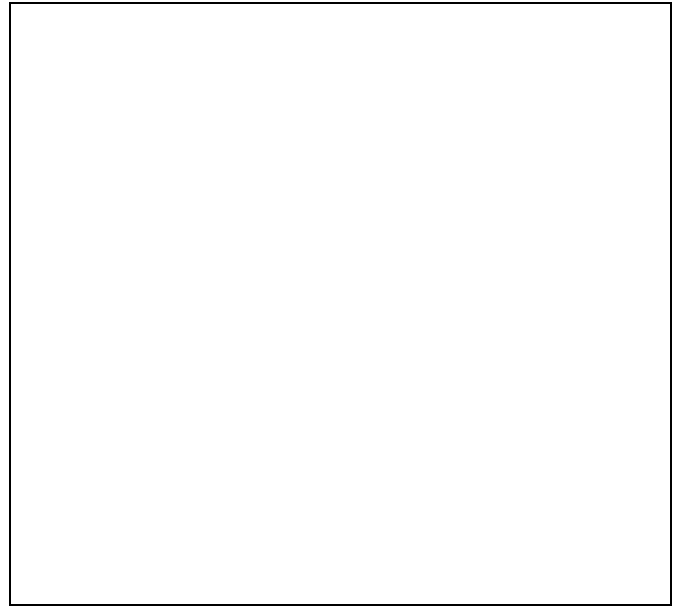

\picplace{8cm}
\caption []{
Residuals of the X-ray fit overlaid with the R CCD image (zoomed image
of 2\arcmin $\times$ 2\arcmin ), with linear X-ray contours. The first
contour is 9 $\times 10^{-7}$ counts s$^{-1}$ arcsec$^{-2}$, and the step
4.1 $\times 10^{-7}$ counts s$^{-1}$ arcsec$^{-2}$.
}
\end{figure}

As the process of subtracting a fitted ellipse implicitly assumes that the
cluster has a perfect elliptical symmetry, and therefore may bias the
significance of the residuals, we checked the above results independently,
with a wavelet analysis of the central part of the X-ray image
(1\arcsec\ pixels not filtered). We used the method described by Slezak et
al. (1994) and looked more specifically at two scales: 5\arcsec\ (HRI PSF)
and 20\arcsec. Results are displayed on Fig. 4.

\begin{figure}
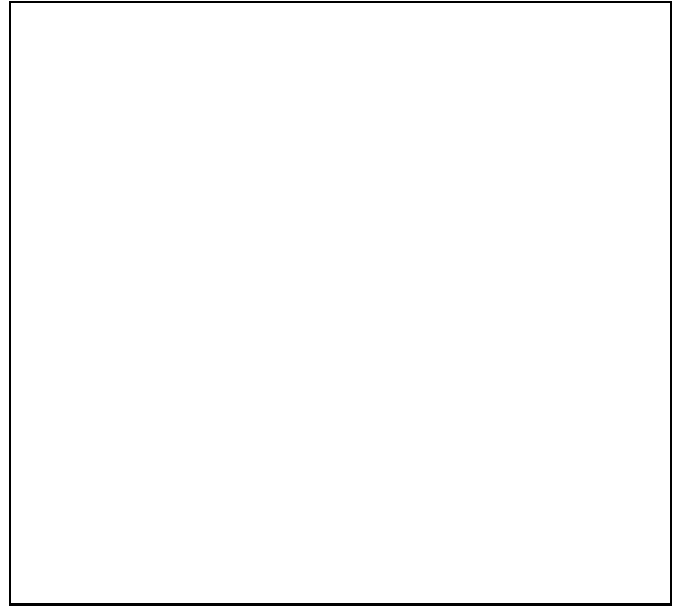

\picplace{8cm}
\caption []{
Wavelet analysis of the central part of the X-ray map. Full line: scales
of 5\arcsec\ (point-like sources); the two blobs P1 and P2 are visible, and
located near the two galaxies labeled as in Paper II.
Dotted line: contributions from scales of 20\arcsec\ which reveal,
in addition to the central region, the diffuse clump S located beyond
the arc. The first contour on both scales is at least 3$\sigma$
above the local fluctuations; X-axis and Y-axis are in decimal degrees.}
\end{figure}

For the smallest scale, the two point-like sources P1 and P2 are detected
again, as well as a strong contribution from the cluster center identified
with the core of the cooling flow and elongated in the
N-S direction. This excess of emission has a direction nearly perpendicular
to the main orientation of the cluster, and the brightest part coincides
with the cD position. However, the small number of photons available does not
allow to assess whether this can be made of two point sources.
In order to check the reality of its elongated shape,
we produced two sub-images selecting
photons according to their arrival time on the detector, half photons in
each image. The N-S elongation remains in both sub-images, and thus can be
considered as real.

The two blobs P1 and P2 deserve also some comments. First, we notice that
the position of P2 is consistent with that of galaxy \#364, a ``normal''
elliptical galaxy at redshift z=0.2334 (Paper II). P2 is by far the
brightest signal, after the
cluster core, observed at this scale over the whole cluster extent. A
possible interpretation could be X-ray emission originating from  hot gas
trapped in a dark matter halo associated to the galaxy \#364. The second
blob on the other side of the cD, P1, is also located near the galaxy
\#314, likely to be a cluster member. But in both cases, the
residual uncertainty of a few arcseconds in the X-ray/optical alignment
leaves room for alternative explanations and the
coincidence with the underlying galaxies could well be fortious. Thus,
another interpretation of  these small scale structures  could relate
them to the presence of the cooling flow. As suggested above, there may
be some inhomogeneities
in the dark matter distribution, and the infalling gas can be trapped in
secondary potential wells such as P1 and P2. Moreover, if the infalling gas
is cooler and denser than the surrounding medium, it has a higher
X-ray emissivity in the ROSAT band and may be easier to detect.
One striking point is the global alignment of P1 and P2 with the main
elongation of the cluster and the nearly perfect alignment with the
central X-ray emission. This kind of features has already been detected in
two well-known cooling flow clusters, namely A85 and A496 (Prestwich et
al. 1995). In these two clusters, some residuals are detected in the
central region of X-ray emission,  possibly unresolved at the ROSAT HRI
resolution, and with typical intensities of 3 to 10 $\times 10^{41}$
ergs s$^{-1}$ and distances to the center of 10--15 kpc. In the case
of A2390, the corresponding values are of the order of $10^{42}$ ergs s$^{-1}$
and 50 kpc, slightly larger. They may reflect the intensity of the cooling
flow, as discussed below.

On scales of 20\arcsec\ the contribution from the cluster center is still
large, but a significant emission is again found some 40\arcsec\ from the
center, definitely confirming the presence of the diffuse component S of
X-ray emission located beyond the giant arc.

\subsection{Optical observations and galaxy distribution}
In order to compare the X-ray map to the light distribution of the cluster,
a photometric catalog of the galaxies was produced from deep images of the
cluster in B and R,  covering a larger field  than the images analysed
in Paper II and best suited to the extent of the X-ray image.
The CCD images were obtained in the course of a systematic
deep imaging of cluster-lenses performed in collaboration with J.A. Tyson
and collaborators, at the Kitt Peak 4 meter telescope in August 1989. Images
were obtained, with the B$_J$ and the R filters,
for which the photometric system is defined in Gullixon et al. (1995).
The details of the
photometric analysis of the cluster will be published elsewhere (Tyson et al.,
in preparation), but we can shortly summarize here the main characteristics
of the observations. The field of view is $8\arcmin \times 8\arcmin$ with a
pixel size of 0.47\arcsec and a seeing on each final image of 1.2\arcsec .
The total exposure time was 2h 20mn in R and 4h
in B$_J$, giving a detection level of $\mu_R = 27.5$ and $\mu_J = 28.3$ at
a $1 \sigma$ level. Photometry of the field was performed using  the standard
FOCAS routines (Valdes et al. 1983). We estimate the
completeness limit of the final catalog to R = 23.8 and B$_J$ = 26, with
more than 1200 objects within this limit. For the purpose of the present
paper, we derived two maps from this catalog,
namely the map of isoluminosity contours and the map with isopleth contours.
Selection criteria were used in order to reduce the contamination by
non-member objects. First, only the objects classified ``galaxy'' by
the FOCAS classification scheme were kept in the catalog. We also cut
the catalog down to its completeness limit ($16.9 < R < 23.8$) and introduced
a color selection
($1.5 < B_J - R < 2.5$, with color indices computed in the B$_J$ isophote)
as the sequence of the ellipticals falls in this
range in the color-magnitude plot. This selection allows a significant
reduction of the statistical
noise introduced by the star contamination. Our final working sample
contains 517 objects. Each map was computed on a regular grid
of points every 10 pixels, with the method developed by Dressler (1980),
{\em i.e.} the radius of the 10th most
distant object from each grid point defines the integration surface.
The accuracy in the determination of the maximum of the map is about 1 grid
point, or 5\arcsec on the final map. A smoothing with a gaussian filter
25 pixels wide was applied, giving a resolution of 12\arcsec, similar in
average to the X-ray map.

If we assume that the X-ray distribution follows the gas density
distribution, we
can also compare this distribution with the galaxy distribution. As seen
in Fig. 5, the optical luminosity distribution  coincides with the X-ray
centroid and  is strongly dominated by the central cD. On the contrary, the
isopleth
map is less convincing in the sense that there is no strong overdensity
of objects near the center, probably because of an antisegregation effect,
{\em i.e.} a default of galaxies near the giant
one, due to cannibalism effects and dynamical evolution in the cluster core.
Moreover, there is a global elongation of the cluster in the direction
$-50^\circ$ from the N-S axis, similar to the X-ray map. The ellipticity
of the luminosity map is typically of $b/a \sim 0.7$, again similar to
the large scale ellipticity of the X-ray. This result can be compared with
the analysis of 5 Abell clusters proposed by Buote and Canizares (1992)
who find, contrary to this cluster, a significant disagreement between
the X-ray ellipticities and the galaxy distribution. Projection effects
on the ellipsoid may be one reason for this discrepancy. The alignment
of the cD along the privileged direction can also be noticed, as it is the
case in many cD clusters.  Finally, one may note that the
galaxy distribution is more clumpy than the X-ray, in particular at larger
distances. Indeed at least two clumps of galaxies located on each side
from the center are clearly visible in Fig. 5, but do not show significant
counter-parts in the X-ray map. They probably correspond to
some subcluster components interacting with the much more evolved main
potential, which was able to accumulate the hot gas since its formation.

\subsection{Existence of the X-ray sub-structure ``S''}
{}From the X-ray emission of the NW sub-structure, we can roughly estimate
its contribution to the total mass. As written above, the X-ray luminosity
is roughly 100 times lower than the total one so this corresponds to
 an associated ``velocity dispersion'' 5 times smaller than the central
one, if we assume that the X-ray luminosity varies as $\sigma^3$
(Edge \& Steward 1991). The mass
is also assumed to vary as $\sigma^2$, which is exact in the case of an
isothermal mass distribution, and can be considered as a first order
approximation otherwise. So the mass associated to the sub-structure is
about 20 times smaller than the cluster mass, a rather standard value
for sub-structures associated to rich clusters of galaxies. The definite
confirmation of its dynamical existence would be to detect a secondary
clump in the velocity histogramme of cluster members. But as shown, its
width would be only 3 to 5 times smaller than the main distribution.
Moreover, from Paper II, the redshift for the main galaxy of the clump
(\# 388, the galaxy near the arc) is 0.231, or a zero radial velocity
with respect to the cD. A possible subcomponent in the histogram
would be centered roughly at the same redshift as the main one!
The velocity histogramme issued from the spectroscopic survey analysed
in Paper II shows a positive kurtosis which could be interpreted as the
addition of a gaussian with a width of 600 km s$^{-1}$ on the main
gaussian, 2000 km s$^{-1}$ wide. Indeed the galaxies in the bin $\pm$
600 km s$^{-1}$ are highly concentrated near the cluster center, in a
region including the clump ``S'', while the galaxies having higher
radial velocity have a smoother distribution over the cluster. This
might be a further evidence for the dynamical reality of the clump,
although it is difficult to spatially separate the two samples.
\begin{figure}
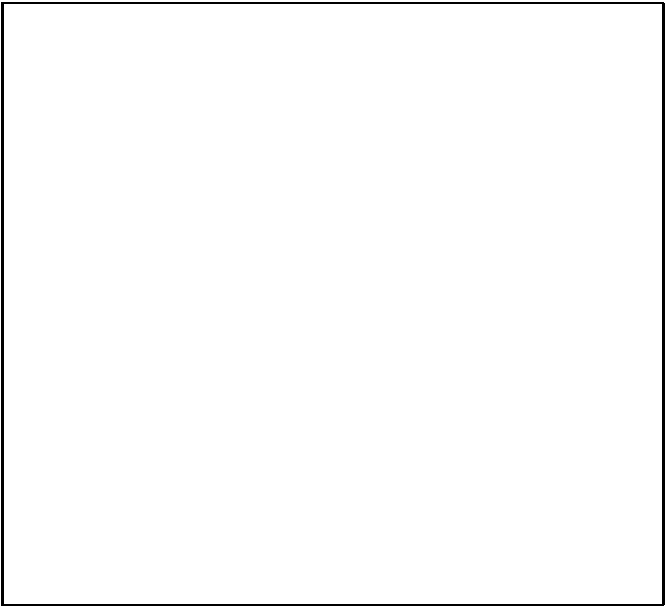

\picplace{8cm}
\caption []{Luminosity density contours overlaid on the R image of A2390.
The map was smoothed to give a typical resolution of 12\arcsec, similar
to the large scale distribution of the X-ray gas.
}
\end{figure}

\section{Analysis of the X-ray flux and profile}
\subsection{X-ray profile}
Because of the significant elongation of the cluster, elliptical
coordinates are the best adapted for the determination of the
brightness profile (Buote and Canizares 1992, 1994). Since the
ellipticity parameters vary with the
radial distance, we adopted mean values corresponding to a region
where  the surface brightness has decreased
by a factor of 2 to 3 from its central value
(i.e. a major axis of $\sim$ 50\arcsec ); this yields $b/a = 0.74$ and
PA = -- 51$^\circ$ (same definition as in Fig. 2).

After the image transformation,
photons were binned into concentric rings and the resulting surface
brightness profile is displayed in Fig. 6. A King model
$$S(R) = S_{0}(1+(R/R_{c})^{2})^{0.5-3\beta} +bkg$$
was fitted to the data, introducing simultaneously to the fit a
deconvolution by the HRI PSF (David et al 1992).
The overall profile is extremely steep and leaves the innermost points in
excess as the signature of the
cooling flow when the fit is extended out to 500\arcsec. A tractable
analytical expression for the profile was obtained by fitting a King
model only up to the point where the profile reaches the background level,
{\em i.e.} a radius of about 250\arcsec. With this restriction the fit
was satisfactory, giving $\beta=0.41$ and $Rc=7.0\arcsec $ for
$0<R<250\arcsec$ and a background value
of $1.15 ~10^{-6}$ counts s$^{-1}$ arcsec$^{-2}$.
To check the validity of our
assumption, we computed the expected 2D surface brightness taking a
3D gas density given by:
$$n(r)= n_{0}(1+(r/Rc)^{2})^{-3/2\beta}\ ~~~ {\rm for} ~~~
0<r<250\arcsec$$
and $$n(r) = 0 ~~~ {\rm for} ~~~ r>250\arcsec$$
using the above values for $\beta, ~Rc$ and the background level.
The modelled surface brightness
convolved by the HRI PSF is shown on Fig. 6 in comparison with the
observed profile and appears to be fully satisfactory. The
corresponding density will be adopted in the rest of the paper.
For this calculation, we have implicitly
assumed that the observed surface brightness is weakly dependent
on the gas temperature,
which is justified, according to the expected high temperature (even in the
cooling flow region) of this very bright cluster, and the poor
temperature sensitivity of the ROSAT HRI above 5 keV. Moreover, this very
uncommon
combination of $\beta$ and $Rc$ should not be  ascribed any physical
meaning, but is only indicative of the presence of cooling flow,
its main advantage  being  to
allow a straightforward derivation of the density profile.

\subsection{Total luminosity}
The  cluster luminosity was calculated integrating counts up to a
maximum radius and corrected with the background value determined
above. Corresponding count-rates were converted into luminosity using
a Mewe-Kaastra thermal model folded with the HRI response provided by
the EXSAS package. Assuming a thermal spectrum of 9 keV (cluster average
temperature), a heavy element abundance of 0.5 in solar units
and taking a galactic neutral hydrogen column density at the
cluster's position of $N_{H}=6.7~10^{20}$ cm$^{-2}$ (Dickey \& Lockman 1990),
we obtain respectively
\begin{eqnarray*}
Li = 5.63\, 10^{44} \, h^{-2} {\rm ergs \ s}^{-1} \quad & {\rm for} \quad
Ri = 250\arcsec \\
Lo = 6.30\, 10^{44} \, h^{-2} {\rm ergs \ s}^{-1} \quad & {\rm for} \quad
Ro = 400\arcsec
\end{eqnarray*}
in the 0.1-2.4 keV band depending on the radius of the integration limit
for the total flux. This is significantly higher than the Einstein
luminosity ($L_X = 3.85 \, 10^{44} \, h^{-2}$ ergs s$^{-1}$ in the
range 0.7--3.5 keV, Ulmer et al. 1986); the difference is too large to
be attributed to the influence of the cooling flow only, although the
ROSAT band is softer than Einstein, and results presumably from an error
in the analysis of the Einstein data (as it is also the case for the very
bright cluster A2163, Elbaz et al. 1995).
Fig. 6 also shows that $Ro$  is
a conservative boundary for the flux integration; however, because of
the absence of profile modelling beyond 250\arcsec\
and the rather high background of the HRI, it is difficult to
estimate how far the cluster emission actually extends, and
consequently apply any correction for the total luminosity.
Normalizing the density profile according to $Li$, we obtain $n_{0}
= 5.7 ~10^{-2}~ \sqrt h$ cm$^{-3}$ which is unusually high.
Using the Temperature - Luminosity relation (Henry \& Arnaud 1991) we derive
a temperature of $\sim $ 9 keV. This is however only indicative, for
the dispersion in the empirical relation is large mostly  because of  possible
cooling flows. Having no observational constraints on  temperature,
we shall adopt this value  as the  mean cluster temperature in the following.

\begin{figure}
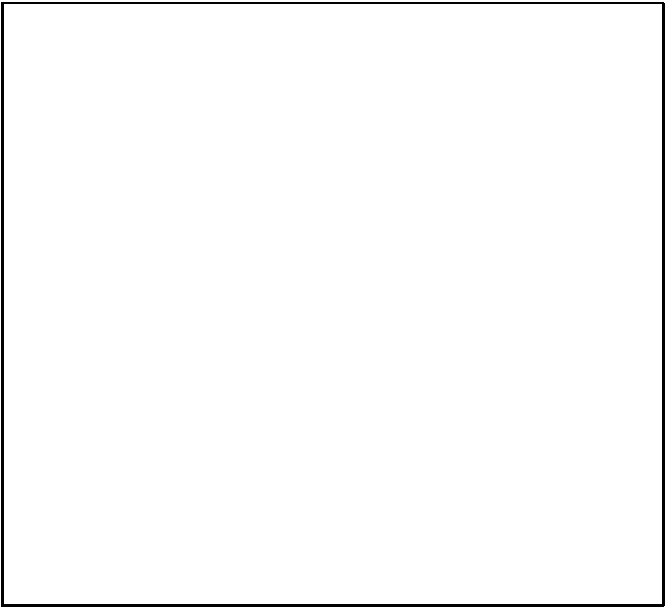

\picplace{8cm}
\caption [] {Surface brightness profile. A bin size of 2.5\arcsec\ was used in
order to have a central annulus of 5\arcsec\ diameter  to fully exploit
the HRI resolution. The solid line is the best fit obtained assuming
a King density profile truncated at R=250\arcsec\ (see text)}
\end{figure}

\subsection{Evidence for a strong cooling flow}
In this section, we propose to estimate the cooling radius and the
mass flow rate in this cluster. Since the electron density scales as
$\sqrt h$ and the time as  $h^{-1}$, these quantities are not
simple functions of $h$ and we give a
detailed calculation in the two cases: $h=0.5$ and $h=1$.
The cooling time as a function of density may be estimated by
(Henry and Henriksen 1986):
$$T_{cool}(yr) = 6~10^{10} \,\left(\frac{T}{10^{8}K}\right)^{1/2}
\, \left(\frac{\rho}{10^{-3}cm^{-3}}\right)^{-1}$$
Using the density profile determined above and assuming again a temperature
of 9 keV, we derive $R_{cool}$ defining the region where
the cooling time is smaller than the age of the universe at the
cluster redshift ($t_{univ}= h^{-1}10^{10}(1+z)^{-3/2}$) . This gives
$R_{cool}=  43\arcsec$
(200 kpc) for $h$=0.5 and   $R_{cool}= 32\arcsec$ (75 kpc) for $h$=1.
Calculating the bolometric cluster luminosity
enclosed within this radius ($L_{cool}$) and assuming a steady state isobaric
gas cooling from 9 keV, we derive the mass flow (Fabian et al. 1991):
$$\dot{M}_{cool} = \frac{2}{5}\frac{\mu m_{p}}{kT}L_{cool}
\sim 880 \, \frac{L_{cool}}{10^{45}} \, \left(\frac{T}{5keV}\right)^{-1}
\, M_{\odot} {\rm yr}^{-1}$$
This yields $\dot{M}_{cool} \sim  850 ~M_{\odot} {\rm yr}^{-1}$ or
$\dot{M}_{cool} \sim  180 ~M_{\odot} {\rm yr}^{-1}$
for $h$ equals 0.5 and 1; the fraction of the total luminosity enclosed
within the cooling radius being $\sim$ 1/4 and 1/5 respectively. It should
be also noticed that if a standard cooling time of $2~10^{10}$ years is
used (e.g. Edge et al. 1994), the mass rates are to be increased by
$\sim$ 30\%.
Although uncertainties are large
(up to a factor of $\sim 2$, Fabian et al. 1991), the derived mass flow
is amongst the highest rates ever observed (Egde et al. 1994). Similar
values can be derived from the analysis of the optical data (Paper IV).

\section{X-ray and lensing constraints on the gravitational potential}
\subsection{Modelling constraints}
Multiple images modelling have successfully demonstrated
that the mass distribution in the central part of clusters of galaxies
is centered on the brightest cluster galaxy (BCG) with an orientation and
ellipticity comparable to the orientation and ellipticity of the
halo of the BCG.
Moreover, the core radius of the mass distribution is found to be very small
(20--50 $h^{-1}$ kpc), not only in the case of cD-type cluster
(MS2137: Mellier et al. 1993; A370 and A2218: Kneib et al. 1993, 1995),
but also in clusters with multiple galaxies in the center such as
Cl0024+17 (Smail et al. 1995b).
These studies show also that the overall matter distribution is clumpy.
Kneib et al. (1993) found that the mass distribution in A370 is bimodal, a
result that has been confirmed by the ROSAT/HRI image of this cluster
(Fort \& Mellier 1994).
In the case of A2218 (Kneib et al. 1995) it is also necessary to add
some other clump of mass on the second brightest galaxy of the cluster
despite no sub-structure is resolved on the deep ROSAT/HRI image.

Previous lensing model of the straight arc in A2390 were done in Paper I
and by Kassiola, Kovner \& Blandford (1992) and Narasimha \&
Chitre (1993). These studies however do not reflect the ROSAT/HRI
observations, in particular the lip-caustic model
of Narasimha \& Chitre (where the center of the mass distribution is very
close to the straight arc) and
the unrealistic fold model and beak-to-beak model with two identical large
masses (see Figures 6 \& 7 of Kassiola et al.).
Their beak-to-beak model with a marginal cluster and two small masses
(their Fig. 8) is probably the closest to the X-ray observations. However the
overall  ellipticity and orientation of their main clump is incorrect, and
their model needs fine tuning to reproduce the straight arc.

The detection of the NW sub-structure ``S''
in X-ray allows us to put strong constraints on the location of the secondary
deflecting mass and its
intensity relative  to the main peak in the construction of the global
gravitational potential.
Therefore the measured distortion and orientation of the straight
arc allows to calibrate the absolute central mass distribution.
The clump acts as a pertuber as shown in Paper I and locally
enhances the convergence and the shear. The shear direction is then
almost straight around the perturber if it has a somewhat smooth
mass distribution (with a core size  of the order of
at least the size of the straight arc -
which is not the case of the prefered model of Kassiola et al.).

To quantify the amount of mass in the central part we have done a simple
mass model taking into account the above considerations, as well as the
geometrical constraints coming from the X-ray luminosity map.
Indeed, the lensing constraints are relatively poor in this cluster,
 since no multiple images are identified; thus, in order to constrain the mass
model, we had to fix some parametres of the mass distribution.
We shall show here that a mass model defined with two clumps of mass
(the main one centered on the cD galaxy and the other one centered on the
``sub-structure") is compatible
with both the X-ray data and the lensed features.
A single clump lens-model cannot reproduce the observed
lensed image unless it is not centered on the cD galaxy and
has a core much larger than the ones found in other cluster-lenses.
Adding a second clump of mass with a core of the size of the straight
arc will generate locally a flat mass distribution
that will have an almost constant convergence and shear direction,
and therefore will produce highly amplified arcs and arclets, straight and
parallel, as expected from the optical observations of Paper I.

The density profile used is the one of Kassiola and Kovner (1993) of the form:
$$
\Sigma(R)=\Sigma_0{R_C\over{\sqrt{R_C^2+R^2}}}
$$
We fixed the ellipticity and orientation of the main clump to be the ones
of the cD halo or the X-ray map at a typical radius of 50\arcsec\,
namely $b/a = 0.7$ and PA = $-49^\circ$.
In order to be consistent with the results found in previous cluster-lenses
(see discussion above) we fixed the core radius of the central clump
to be $R_C \approx$ 13\arcsec\ (30 $h^{-1}$ kpc).
For the secondary clump we assume for simplicity
a circular shape and a core radius of the
same order (7--12\arcsec ). Therefore there are only two unknown
parameters, namely the
central mass density of both clumps. These are easily constrained by the
shape and orientation of the straight arc as well as the shear direction
in that region. The straight arc puts strong upper limits on the overall
mass as it has no counter image candidate and thus limit the extension
of the critical line up to its position. It also gives a lower limit
of the mass as it needs to be well magnified for an object at redshift
0.913 (Paper I).
Table 2 gives the fiducial mass model which is compatible with the straight
arc and the shear around it. Figure 7 shows the predicted shear map
around the cluster center.

Quantitatively, the projected mass we find in a circle of radius 38\arcsec\
or 88 $h^{-1}$ kpc,
corresponding to the distance of the arc is:
M$_{tot}$($<R_{arc}$) = 0.8$\pm$0.1 10$^{14}$ $h^{-1}$M$_\odot$.
The integrated luminosity inside the same surface has been computed in
the photometric catalogue including only the ``galaxies'' but with no
color selection. We also added a k-correction of 0.242. Finally, the
total luminosity is L$_{Rtot}$($<R_{arc}$) = 3.0$\pm$0.1 10$^{11}$
$h^{-2}$L$_\odot$, and
the corresponding mass-to-light ratio is M/L$_R \simeq 260 \pm 40 \, h$
(M/L)$_\odot$. The velocity dispersion inferred from the lensing model
(Table 2) is about twice smaller as the value published in Paper II,
but is consistent with the lower $\sigma_{v}$ fond within the
central region of the cluster  and discussed in Sec. 2.4.

\begin{table*}
\caption[ ]{Fiducial parameters of the two projected potential found for
the best model of the arc.
}
\begin{flushleft}
\begin{tabular}{lllllllll}
\noalign{\smallskip}
\hline
\noalign{\smallskip}
     & x$_c$ & y$_c$ & $\epsilon$  & b/a & PA & R$_C$ &
 $\sigma_\infty$
 & $M_{tot}(<R_{arc})$\\
 & (\arcsec ) & (\arcsec )& ${a^2-b^2\over a^2+b^2}$ & & & (\arcsec )
 & km s$^{-1}$ & $10^{14}\  h^{-1}\ M_\odot$ \\
\noalign{\smallskip}
\hline
\noalign{\smallskip}
cD clump &  0. & 0. & 0.3 & 0.71  & 49.2$^o$ & 12$\pm 5$ & 950$\pm 100$ &
0.8$\pm 0.1$ \\
sub-clump & 42$\pm 2$ & 10$\pm 5$ & ---
& --- & ---  & 7--12 & 420--500 & ---\\
\noalign{\smallskip}
\hline
\end{tabular}
\end{flushleft}
\end{table*}

\begin{figure}
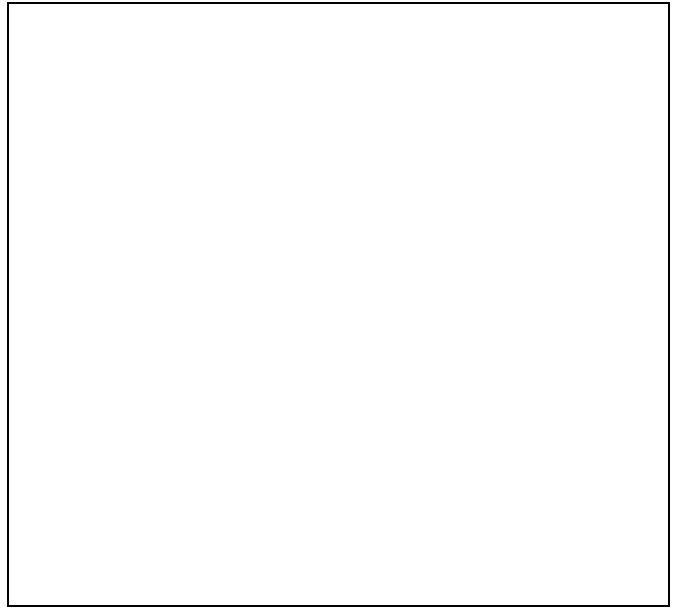

\picplace{8cm}
\caption [] {
Results of the lensing model overlaid on the
optical image of A2390. The contours represent the mass distribution and
the sticks the direction and amplitude of the shear at redshift 0.913,
{\em i.e.} the giant arc redshift.}
\end{figure}

\subsection{Dark matter and cooling flow}
In the rest of the paper, we assume a Hubble
constant of 50 km/s/Mpc for the sake of clarity in the comparison with other
observations and numerical simulations.

It is usually impossible to infer constraints on
the temperature of clusters of galaxies when no
X-ray spectral information is available, and one
is usually left with the temperature estimate provided
by the L$_X$-T relation (cf Sect. 3.2). But in the case
of A2390, the presence of the giant gravitational arc provides
an elegant alternative to classical temperature
measurements, in the sense that it strongly constrains
the shape and the strength of the underlying potential
(at least for the central part). In the hypothesis of hydrostatic equilibrium
(which is justified, since cooling flows are supposed to be highly
subsonic), the derivation is straightforward.\\
The 3D mass distribution derived from the projected mass density used in
the lensing simulation is:
$$m(r) = m_{0} f(x)$$
where
$$x = r/R_C, ~~~~f(x)=x-\arctan (x),~~~~~ m_{0} =  4\pi \rho_{0}
R_C^{3}$$
Replacing by the numerical values, i.e. $R_C$ = 13\arcsec\
and the total projected mass enclosed within $R_{arc} = 38\arcsec $
(0.8$ ~10^{14}~h^{-1}~M_{\odot}$), we find $m_{0} =
2.35~10^{13}~h^{-1}~M_{\odot}$. Then, assuming that the potential
is well represented by the lensing mass profile extrapolated
up to the radius  where  diffuse gas emission is observed, we can
solve the hydrostatic equilibrium:
$$\frac{1}{\rho}\frac{dP}{dr} = -\frac{Gm_{o}}{r^{2}}f(x)$$
assuming an ideal gas
$$\rho(r) = m_{p}n(r)\mu~~~~ P~=~nkT$$
The equation was integrated taking the gas density, $n(r)$, derived in
Sect. 3.1 and an outside temperature of  9 keV,
as boundary condition.
\begin{figure}
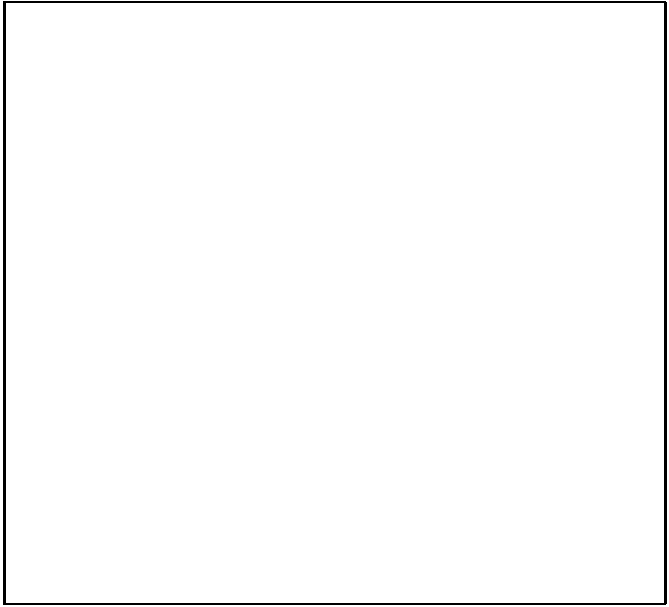

\picplace{8cm}
\caption [] {Full line: cluster profile temperature calculated assuming the
gravitational potential  derived  from the lensing analysis (indepedent of
$Ho$). The dotted and dashed lines correspond respectively to models where
either the gravitating mass has been divided by 2 or the lensing radius
multiplied by 2.}
\end{figure}

The corresponding temperature profile is displayed  on Fig. 8 (solid line),
showing a significant temperature drop toward the center as well as a
negative temperature gradient beyond  $\sim R_{arc}$. Although this is
perfectly plausible with the cooling flow picture, it is difficult to
estimate the uncertainty on the profile. Assuming a higher outside
temperature flattens the outer temperature gradient but leaves $T_{max}$
almost unchanged. On the other hand, one can certainly explore
the influence of the lensing parameters. Two alternatives modifying
the lensing constraints have been investigated:
(1) $~m_{o} ~ \rightarrow ~ m_{o}/2$   and
(2) $R_C  ~ \rightarrow ~ 2 R_C$;
the corresponding temperature profiles are also  presented on Fig. 8.
The resulting profiles are also consistent with a cooling
flow scenario, but show significantly different inner temperatures.
The values chosen for these alternatives are, however, {\em excluded} from
the lensing analysis and
it clearly appears that it would be extremely difficult to constrain
further the temperature profile of this distant cluster.
Technically, it is very unlikely that present day spatially
resolved X-ray spectroscopy may discriminate between the different
possibilities (ASCA has only a spatial resolution of $\sim 3'$). But
the forthcoming XMM, with a resolution of $\sim 30\arcsec $ may be able to
verify these results. Anyway,  a single observation with ASCA should
provide a good estimate of the cluster mean temperature, which is one of
the main unknown of the problem.
\begin{figure}
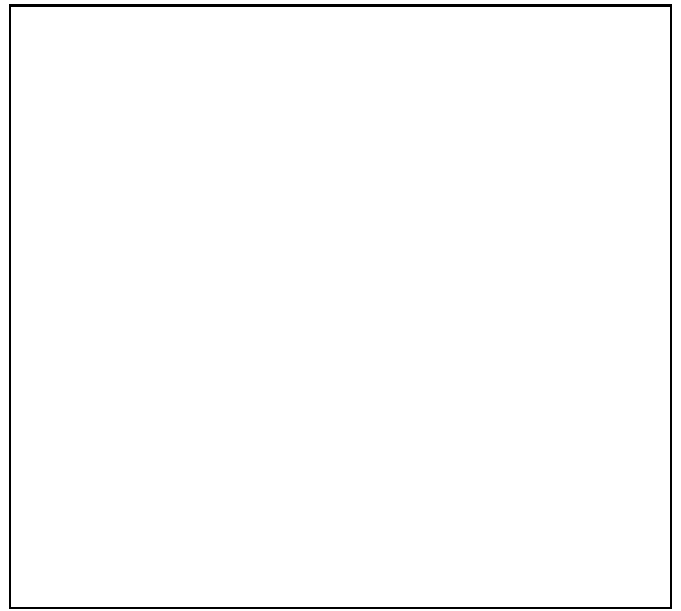

\picplace{8cm}
\caption [] {Integrated total mass profile (full line) and
gas mass profile (dotted line). At $R = R_{arc} = 38\arcsec $,
the gas fraction represents only $\sim$ 1/35 of the total mass.
(computed for $Ho$=50, $M_{tot}$ scales as $h^{-1}$, $M_{gas}$ as  $h^{-2.5}$)}
\end{figure}

The other source of uncertainty is the potential shape at large radial
distances. In the hypothesis where the analytical expression of the
lensing potential holds up to the cluster periphery, we have computed the
integrated mass (gas mass and total mass) as a function of radius (Fig. 9).
At the radial distance of  the giant arc, the gas represents
only $\sim 1/35$
of the total mass, whereas at 250\arcsec\  this fraction reaches 1/12.
This last value is in agreement with what was found previously
in rich clusters (B\"ohringer 1994),
but needs to be confirmed  with a proper definition of the cluster
potential beyond the giant arc (e.g. using weak shear methods).
On Fig. 9, it is also conspicuous that the dark matter is
more concentrated than the gas distribution, despite
the presence of the cooling flow.

With the above temperature profile a direct analytical expression
of the mass flow rate can be derived, using the equation of
energy conservation:
$$\rho v \frac{d}{dr}\left[\frac{v^{2}}{2}+\frac{5}{2}\frac{kT}{\mu
m_{p}}+\Phi\right] = n^{2}\Lambda T^{1/2}$$

Taking the lensing gravitational potential,
the derived temperature profile, a cooling constant
$\Lambda = 1.63~10^{-27}$ $cgs$ (corresponding
to the cooling time used above) and neglecting the quadratic
velocity term, we find a mass flow rate:
$$\frac{dM}{dt} = 4 \pi \ r^{2} \rho v$$
of $\sim 710 ~M_{\odot}$ yr$^{-1}$ for a cooling radius of
43\arcsec\, which is in very good agreement with
the estimate obtained in Sect. 3.3 and confirms the presence of a massive
cooling flow in the center of A2390.

\section{Discussion and conclusions}
Throughout this study, Abell 2390 appears to be one of the most
distant cluster of galaxies for which a detailed analysis enabled the
discovery of an unrevalled collection of remarkable properties: high velocity
dispersion, presence of a peculiar gravitational arc, preferential direction
strongly underlined both by optical and X-ray data, high X-ray luminosity,
massive cooling flow, presence of sub-structures.
The purpose of this paper was the analysis of a high resolution X-ray image
in close connection with optical informations. Thanks to the large number
of X-ray photons available, it was possible to take advantage of the maximal
resolution provided by the ROSAT HRI and study the core of this cluster
down to scales of 5\arcsec\ i.e.  $\sim 12 ~h^{-1}$ kpc. The combination of
X-ray and optical observations allowed to bring a coherent picture of the
cluster under the very simple hypotheses of hydrostatic equilibrium
and that ``light traces mass", at least in its ellipticity and its
orientation.

Two main new results have to be highlighted, for they have numerous
consequences as to the dynamical state, and thus the formation processes of
this  massive cluster as suggested by various numerical simulations: (i) a
high X-ray luminosity partly caused by a strong cooling flow
($\sim 250 ~h^{-2}$ M$_\odot$ yr$^{-1}$) and having an elliptical morphology
on large scales. (ii) The presence of a diffuse sub-structure playing a
key role in the modelling of the lensing constraints. We now briefly
discuss our findings.

In addition to the sub-structure which is best seen on scales of the
order of 20\arcsec, two point-like contributions
were detected as well as a strong emission from
the cluster core. The latter is significantly elongated in a direction
perpendicular to the cluster main orientation, however not
spatially resolved in width in the HRI data. It is interesting to notice that
the inner core of A2218 seems to present the same ``counter-alignment"
property (Kneib et al 1995). Apart from this, it is remarkable that all the
sub-structures, together with the filament observed in the optical and
the giant arc are well aligned along the direction defined by the cD main
axis, which is also suggested by the galaxy distribution on
larger scales. The overall cluster X-ray morphology is moreover strongly
elliptical, with the major axis also coinciding with the cD's, even though
the ellipticity parameters are somewhat varying with radial distance.
Such alignments - at least between X-ray shape and cD orientation - seem
to be quite a common property of relatively  bright X-ray clusters
(e.g. Edge et al. 1994; Pierre et al. 1994) and are probably related to
the way they formed. Indeed, recent 3D hydrodynamical simulations
(Katz \& White 1993),
leading to the formation of a Virgo type cluster as to the final mass (we find
for A2390 a mass of $\sim 10^{15}$ M$_\odot$ within 1.2 Mpc) in
a CDM model with $\Omega _{b} =0.09$, shows that filaments begin to form
by z $\sim$ 0.4. Dark matter halos continue to ``condense" out of the
filaments; material can flow large distance (over 11 Mpc) along the
filaments to reach the forming cluster, while the material that falls into
the cluster without
flowing down a filament is of much more local origin. This gives the
general impression that cluster formation occurs by flow along a set of
intersecting filaments rather than by quasi spherical collapse.
The fact that A2390 presents strong alignment properties supports
very well this picture.
Katz \& White simulations show moreover that the intra-cluster medium is
far from isothermal: in addition to a central cooling flow, the temperature
falls with radius throughout the virialized region of the cluster. This is
qualitatively in perfect agreement with our X-ray analysis if we assume
that the potential shape, constrained by the lensing up to $\sim$ 40\arcsec\
holds at much larger radial distances.

A recent X-ray analysis of the  simulated Katz \& White clusters
by Buote \& Tsai (1995)
provides interesting assessments  as to the actual validity  of such a
study where one attempts to derive the dark matter profile from the X-ray
luminosity distribution under the hypothesis of hydrostatic equilibrium.
Generally speaking, in order to generate the same ellipticity of the potential
at a given distance, a cluster having a negative temperature gradient for the
X-ray emitting gas will have to be more elongated than a cluster  having an
isothermal gas. They find that conclusions regarding the shape of the dark
matter are not overly sensitive to the temperature gradient of the gas.
In particular, for redshifts between 0.25 and 0.13 there is an
excellent agreement between the X-ray and true dark matter shapes
provided that the true inclination of the cluster is taken
into account; otherwise the deviation from the true ellipticity increases
considerably. This is undoubtedly one of the major limitations of the
present study but, owing to the internal consistency of our results and the
apparent strong elongation of the cluster, we
are confident that the symmetry axis of the cluster should be within
$\pm 30^{o}$ from the plane perpendicular to the line of sight.
The presence of the giant arc in A2390 allowed to infer
the shape of the dark matter distribution with an independent method.
Surprisingly, all the distributions (X-ray gas, dark matter, galaxies)
present the similar ellipticities, reinforcing the conclusions discussed here.

Finally, the extended structure discovered in the X-ray emission
agrees well with
what was expected for producing the exceptional shape of the giant arc,
namely the presence of a secondary deflecting mass located beyond the arc
and aligned with the cD-arc axis as already suggested in Paper I. On the
qualitative point of view, it is
worth underlying the analogy between the distribution of matter that
we observed around the arc and Kassiola et al. (1992) ``prefered model" (e.g.
their Fig. 8). Although their numerical solution does not match exactly
our findings, their general idea should be regarded as valuable. Such a
distribution of matter predicts,
however, that the giant arc should be made of two galaxy images, a fact
which is corroborated (as noticed by  Kassiola et al.) by the infrared
observations of Smail et al. (1993). Further HST observations will most
probably disentangle the uncertainties on the morphological shape of the
source of the giant arc. It is worth noting that similar X-ray/optical
studies of cluster-lenses have already brought some indications on the
dynamical state of rich clusters of galaxies. In particular, in the
clusters A370 and A2218, the lens modellings have revealed significant
departures from the virialised state, in agreement with the irregularities
of the X-ray maps (Kneib et al. 1995). On the contrary in the case of A2390,
the gas distribution is more regular, and dominated by a massive cooling
flow a phenomenon which can appear only if the dynamical secular evolution
is far from violent, and is destroyed during merger phases for example.
This is a valid argument to understand why this cluster appear rather
regular in its X-ray distribution.

In this paper, we have only discussed the matter distribution of the cluster
on its central regions. More extended analysis will benefit from a
weak shear approach, {\em i.e.} the large scale statistical analysis
of the gravitational deformations. Recent theoretical works have proposed
different methods to invert the shear map to recover the mass distribution
(Kaiser \& Squires 1993, Seitz \& Schneider 1995).
Detection of the weak deformation on the faint background population
at the periphery of the clusters (Bonnet et al. 1994; Fahlman et al. 1994;
Smail et al. 1995c) are very promising, as they should provide strong
constraints on the slope of the mass profile.
The mass estimated by this technique seems presently quite large
(Bonnet et al. 1994; Fahlman et al. 1994), but the uncertainties are
large too and critically dependent on the quality of the observations.
The future comparison between these mass maps and the mass derived from
large scale X-ray distribution such as those from ROSAT/PSPC data will
be important to confirm the apparent increase of dark matter outside
the central regions of rich clusters of galaxies.
If these results are confirmed, it could have important cosmological
implications.

\acknowledgements
We wish to thank Y. Mellier and R. Pell\'o for their constant interest
in this work and all their comments on this paper,  B. Fort,
G. Mathez and A. Blanchard for interesting discussions on clusters of
galaxies, lensing and gas dynamics and M. Arnaud for the use of her plasma
codes. We are also grateful to J.L. Starck for detailed informations on the
wavelet analysis and the use of his package.
JPK is grateful for financial grant from the EC HCM Network CHRX-CT92-0044.
This work was supported by the French Centre National de la
Recherche Scientifique (INSU) and the Groupe de Recherche ``Cosmologie''.


\begin{thebibliography}{}
\bibitem{} Allen S.W., Fabian A.C., Kneib, J.P., 1995, submitted
\bibitem{} B\"ohringer H., 1994, in "Cosmological apsects of clusters of
           galaxies", p 123, ed. Seitter, NATO ASI, Kluwer
\bibitem{} Bonnet H., Mellier Y., Fort B., 1994, ApJ 427, L83
\bibitem{} Buote D. A., Canizares C. R., 1992, ApJ 400, 385
\bibitem{} Buote D. A., Canizares C. R., 1994, ApJ 427, 86
\bibitem{} Buote D. A., Tsai J. C., 1995, ApJ 439, 29
\bibitem{} Daines S.J., Jones C., Forman W., Tyson J.A., 1995, preprint
\bibitem{} David L.P., Harden F.R., Kearns K.E., Zombeck M.V., 1992, U.S.
           {\em ROSAT} Science Data Center/ SAO, {\em The ROSAT High
           Resolution Imager}
\bibitem{} Dickey J.M., Lockman F.J., 1990, ARAA 28, 215
\bibitem{} Dressler A., 1980, ApJS 42, 565
\bibitem{} Edge A. C., Stewart G.C., 1991, MNRAS 252, 428
\bibitem{} Edge A. C., Fabian A. C., Allen S. W., Crawford C. S., White D. A.,
           B\"ohringer H., Voges W., 1994, MNRAS 270, L1-L5
\bibitem{} Elbaz D., Arnaud M., B\"ohringer H., 1995, A\&A 293, 337
\bibitem{} Fabian A.C., Nulsen P.E.J., Canizares C.R., 1991, A\&AR 2, 191
\bibitem{} Fahlman, G. G., Kaiser, N., Squires, G. and Woods, D., 1994,
           ApJ, 437, 56
\bibitem{} Fort B., Mellier Y., 1994, A\&AR 5, 239
\bibitem{} Gullixon C.A., Boeshaar P.C., Tyson J.A., Seitzer P., 1995,
           ApJS 99, 281
\bibitem{} Henry J.P., Arnaud K., 1991, ApJ 372, 410
\bibitem{} Henry J.P., Henriksen M.J., 1986, ApJ 301, 698
\bibitem{} Jedrzejewski R., 1987, MNRAS 226, 747
\bibitem{} Kaiser N., Squires, G., 1993, ApJ 404, 441
\bibitem{} Kassiola A., Kovner I., Blandford R.D., 1992, ApJ 396, 10
\bibitem{} Kassiola A., Kovner I., 1993, ApJ 417, 450
\bibitem{} Katz N. White S.D.M., 1993, ApJ 412, 455
\bibitem{} Kneib J.P., Mellier Y., Fort B., Mathez G., 1993, A\&A 273, 367
\bibitem{} Kneib J.P., Mellier Y., Pell\'o R., Miralda-Escud\'e J., Le
           Borgne J.F., B\"ohringer H., Picat J.P., 1995, A\&A in press
\bibitem{} Le Borgne J.F., Mathez G., Mellier Y., Pell\'o R., Sanahuja B.,
           Soucail G.,  1991, A\&AS 88, 133 (Paper II)
\bibitem{} Loeb A., Mao S., 1994, ApJ 435, L109
\bibitem{} Mc Millian S.L.W., Kowalski M.P., Ulmer M.P., 1989, ApJS 70, 723
\bibitem{} Mellier Y., Fort B., Kneib J.P., 1993, ApJ 407, 33
\bibitem{} Merrit D., 1987, ApJ 313, 121
\bibitem{} Merrit D., Saha, P., 1994, ApJ 409, 75
\bibitem{} Merrit D., Tremblay B., 1994, AJ 108, 514
\bibitem{} Miralda-Escud\'e J., Babul, A., 1995, preprint
\bibitem{} Narashima, Chitre, 1993, J. Ast. Astronomy 14, 121
\bibitem{} Pell\'o R., Le Borgne J.F., Soucail G., Mellier Y., Sanahuja B.,
           1991, ApJ 366, 405 (Paper I)
\bibitem{} Pierre M., Soucail G., B\"ohringer H., Sauvageot J.-L., 1994,
	   A\&A 289, L37
\bibitem{} Prestwich A.H., Guimond S.J., Luginbuhl C.B., Marshall J. 1995,
           ApJ 438, L71
\bibitem{} Seitz C., Schneider, P., 1995, A\&A, in press.
\bibitem{} Slezak E., Durret F., Gerbal D., 1994, ApJ 108, 1996
\bibitem{} Smail I., Ellis R., Arag\'on-Salamanca A., Soucail G.,
           Mellier Y., Giraud E., 1993, MNRAS, 263, 628
\bibitem{} Smail I., Hogg D.W., Blandford R., Cohen J.G., Edge A.C.,
           Djorgovski S.G, 1995a, MNRAS submitted
\bibitem{} Smail I., Dressler A., Ellis R.S.,  Kneib J.P., Couch W.J.,
           Sharples R.M., Oemler A., Butcher H.R., 1995b, in press.
\bibitem{} Smail I., Ellis R S., Fitchett M.J., Edge A C., 1995c, MNRAS
           273, 277
\bibitem{} Soucail G., Arnouts S., Le Borgne J.F., Pell\'o R.,
           Fraix-Burnet D., in preparation (Paper IV)
\bibitem{} Ulmer M.P., Kowalsky M.P., Cruddace R.G., 1986, ApJ 303, 162
\bibitem{} Valdes F., Tyson J.A., Jarvis J.F., 1983, ApJ 271, 431
\end{thebibliography}
\end{document}